\documentclass[sigconf, nonacm, pdfa]{acmart}

\newcommand\vldbdoi{XX.XX/XXX.XX}
\newcommand\vldbpages{XXX-XXX}
\newcommand\vldbvolume{19}
\newcommand\vldbissue{12}
\newcommand\vldbyear{2026}
\newcommand\vldbauthors{\authors}
\newcommand\vldbtitle{\shorttitle} 
\newcommand\vldbavailabilityurl{https://github.com/SolidLao/GenDB}
\newcommand\vldbpagestyle{empty} 

\usepackage{tikz}
\newcommand{\circnum}[1]{%
\tikz[baseline=(char.base)]{
    \node[draw, circle, inner sep=1pt] (char) {\small #1};
}}

\usepackage[a-2b]{pdfx}

\begin{document}
\title{Demonstrating GenDB: Instance-Optimized and Customized Query Processing Code Generation via LLM Agents}


\author{Jiale Lao}
\orcid{0009-0003-1144-5152}
\affiliation{%
  \institution{Cornell University}
}
\email{jiale@cs.cornell.edu}

\author{Immanuel Trummer}
\orcid{0000-0002-1579-3221}
\affiliation{%
  \institution{Cornell University}
}
\email{itrummer@cornell.edu}

\begin{abstract}
Traditional query processing engines require continuous development and extensions to support new techniques and user requirements, and in some cases, entirely new systems must be built from scratch. However, these engines are difficult to extend due to their internal complexity, and building new systems demands significant engineering effort and cost. To address this, we demonstrate GenDB, a generative query engine that shifts query processing from manually engineered systems to query processing code generation driven by Large Language Models (LLMs). An early prototype of GenDB uses LLM agents to generate instance-optimized query execution code tailored to specific data, workloads, and hardware resources. This prototype suits offline code generation for repetitive, templated queries, since the upfront generation cost amortizes over many executions and correctness can be ensured through extensive fuzz testing and manual inspection. For ad-hoc queries, GenDB can work with a traditional DBMS in a hybrid architecture: the DBMS handles one-off queries, while GenDB speeds up frequent SQL templates. We also discuss the current limitations of GenDB, such as high generation cost and the lack of correctness guarantees, to inform future extensions.  Our demonstration allows users to (1) visually and interactively explore how GenDB analyzes workloads, profiles hardware resources and underlying data, produces query plans, generates code based on them, and finally uses an optimizer to iteratively achieve a correct and efficient implementation; (2) use visual inspection and analysis to gain qualitative insights into why GenDB produces code that achieves significantly better performance than state-of-the-art query engines on two benchmarks: TPC-H and a newly constructed benchmark designed to reduce potential data leakage from LLM training data; and (3) upload their own data and queries to explore GenDB with different LLMs and query patterns. A live demo is available at {\color{blue} \url{https://solidlao.github.io/GenDB/demo/}}.
\end{abstract}

\maketitle

\pagestyle{\vldbpagestyle}
\begingroup\small\noindent\raggedright\textbf{PVLDB Reference Format:}\\
\vldbauthors. \vldbtitle. PVLDB, \vldbvolume(\vldbissue): \vldbpages, \vldbyear.\\
\href{https://doi.org/\vldbdoi}{doi:\vldbdoi}
\endgroup
\begingroup
\renewcommand\thefootnote{}\footnote{\noindent
This work is licensed under the Creative Commons BY-NC-ND 4.0 International License. Visit \url{https://creativecommons.org/licenses/by-nc-nd/4.0/} to view a copy of this license. For any use beyond those covered by this license, obtain permission by emailing \href{mailto:info@vldb.org}{info@vldb.org}. Copyright is held by the owner/author(s). Publication rights licensed to the VLDB Endowment. \\
\raggedright Proceedings of the VLDB Endowment, Vol. \vldbvolume, No. \vldbissue\ %
ISSN 2150-8097. \\
\href{https://doi.org/\vldbdoi}{doi:\vldbdoi} \\
}\addtocounter{footnote}{-1}\endgroup

\ifdefempty{\vldbavailabilityurl}{}{
\vspace{.3cm}
\begingroup\small\noindent\raggedright\textbf{PVLDB Artifact Availability:}\\
The source code, data, and/or other artifacts have been made available at \url{\vldbavailabilityurl}.
\endgroup
}

\section{Introduction}


Supporting new techniques and user requirements demands continuous development of existing query processing engines, or in some cases, building entirely new systems. However, the field evolves rapidly, and this paradigm often fails to keep pace: extending existing engines is difficult due to their internal complexity, and building new systems requires substantial engineering effort and cost. For rapidly emerging techniques, should we continue extending existing systems, or design new ones? \textbf{Or is there another option?}

To address this, we demonstrate an early prototype of GenDB that uses LLM agents to generate instance-optimized and customized query processing code tailored to specific data, workloads, and hardware. This prototype suits offline code generation for repetitive, templated queries common in industrial workloads~\cite{redset-vldb}, since the upfront generation cost amortizes over many executions. It uses runtime feedback to iteratively refine the generated code until it outperforms existing query engines or exhausts its budget, and uses fuzz testing to verify that the code returns consistent results across different SQL template predicates and testing databases.



This demonstration lets visitors experience GenDB in action. Visitors will be able to visually and interactively explore how GenDB uses a multi-agent workflow to analyze workloads, profile hardware and underlying data, produce query plans, generate code based on them, and finally use an optimizer to iteratively refine the code. The behavior of each agent is not a black box; each agent's observations, reasoning, and decisions are visualized and can be explored interactively in detail. Such visual inspection and analysis, from high-level optimization strategies to low-level implementation details, also help visitors gain qualitative insights into why GenDB produces code that outperforms state-of-the-art query engines. In addition to the two pre-configured benchmarks, visitors also have the flexibility to upload their own data and queries for experiments.

\section{System Overview}

\begin{figure}[htbp]
    \centering
    \includegraphics[width=0.85\linewidth]{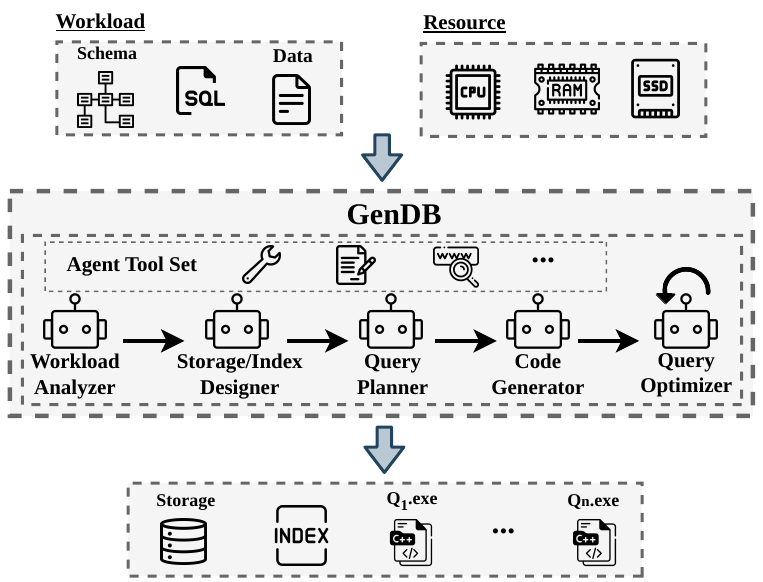}

    \caption{System Overview of GenDB}

    \label{fig:gendb_overview}
\end{figure}

Figure~\ref{fig:gendb_overview} shows an overview of GenDB. It takes as input the schema, SQL queries, database, and available resources. GenDB generates instance-optimized and customized database storage structures, indexes, and one executable file per SQL template. The current prototype targets offline code generation for repetitive and templated queries that are common in industrial workloads~\cite{redset-vldb}, since the upfront generation cost amortizes over many executions.


GenDB is an LLM-powered agentic system that decomposes the complex end-to-end query processing and optimization task into a sequence of smaller and well-defined steps, where each step is handled by a dedicated LLM agent.
First, the Workload Analyzer analyzes the hardware resources, schema, SQL queries, and underlying data to generate structured workload characteristics (e.g., storage type and SIMD support, table and column statistics, join patterns, and filter selectivity). 
Second, the Storage/Index Designer designs and implements the code that transforms the original data formats into optimized storage structures (e.g., using columnar storage with encoding and compression for OLAP workloads) and builds the corresponding indexes. Third, the Query Planner generates an efficient execution plan based on the workload analysis and the storage designs. Each physical operator implementation is resource-aware. 
For example, the aggregation operator adapts its strategy to the hardware cache hierarchy: for a GROUP BY over a low-cardinality column like nation (25 distinct keys), it uses a direct flat array that fits entirely in L1 cache with zero hashing overhead, whereas for millions of distinct groups it switches to thread-local hash tables sized to fit each core's private cache, merging results in parallel only after the scan completes. Fourth, the Code Generator implements the execution plan using appropriate programming languages (e.g., C++ for high performance) and applies system-level optimizations (e.g., compiler optimizations). 
The generated code is executed on the database to measure both correctness and efficiency. 
GenDB validates code correctness by comparing the generated query results with those produced by a traditional database system (available because current prototype targets offline code generation for repetitive and templated queries). Fuzz testing can be used to verify that the code returns consistent results across different SQL template predicates and testing databases.
Fifth, the Query Optimizer uses the feedback to iteratively refine both the execution plan and the code implementation to improve performance. 
After exhausting the user-configured optimization budgets or reaching the user-specified targets, GenDB outputs the instance-optimized storage layouts and one executable file per SQL template.

Each agent in GenDB uses an LLM as its core reasoning component to plan and automatically invoke tools to complete the assigned tasks. Each agent is equipped with built-in tools, including file operations such as reading, writing, and editing files; terminal access for executing commands and scripts; and web-related tools such as web search and web content retrieval. This tool-use capability allows agents to complete complex tasks. For example, if a query contains multi-table joins, an agent can generate a sampling program to evaluate candidate join orders, measure their performance, and refine the code to use the empirically best join order.

\begin{figure}[bp]
    \centering
    \includegraphics[width=1\linewidth]{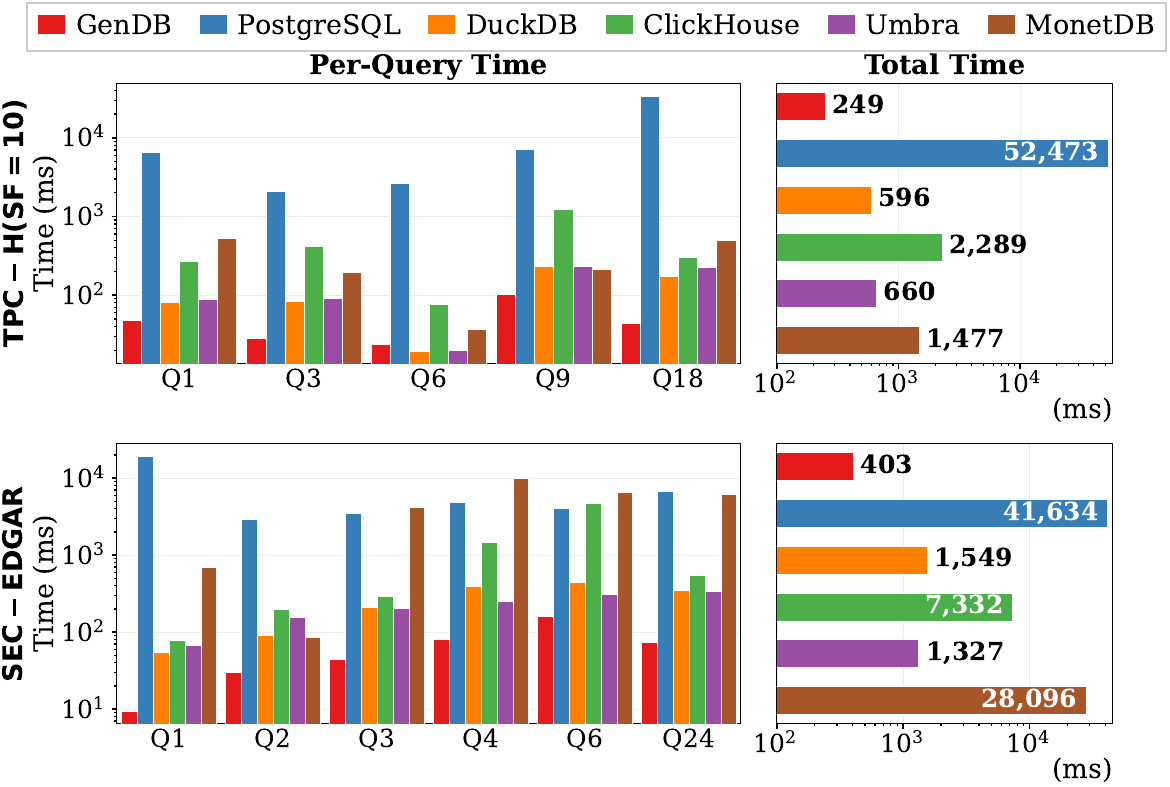}

    \caption{Performance Comparison}

    \label{fig:perf_comparison}
\end{figure}

\section{Extract of Experimental Evaluation}

Due to space limitations, we present a small extract of our experimental evaluation, with more details available in~\cite{lao2026gendbgenerationqueryprocessing}. 

\noindent \textbf{Settings.} GenDB is implemented in JavaScript and employs Claude Agent to build the agentic workflow. All experiments are conducted on an Ubuntu server with two Intel Xeon Gold 5218 CPUs and 384 GB of RAM. We ensure that the entire database is cached in memory to report the execution time of hot runs, following the same setting as in~\cite{leis2015good}. We compare GenDB with state-of-the-art engines: DuckDB, ClickHouse, Umbra, MonetDB, and PostgreSQL, and ensure that all systems use comparable hardware resources (e.g., fully parallelized execution). For fairness, we report each competing engine's best performance across two configurations: (1) using only its automatically created indexes (e.g., the zonemaps OLAP engines create by default), and (2) supplementing these system-created indexes with additional indexes recommended by GenDB. We use (1) TPC-H with scale factor 10, and (2) a newly constructed benchmark, SEC-EDGAR, to reduce potential data leakage, since TPC-H is well studied and represented in LLM training data. SEC-EDGAR uses a dataset that has rarely been used for database benchmarking, and its queries are randomly generated using SQLSmith~\cite{sqlsmith}.

\begin{figure*}[htbp]
    \centering
    \includegraphics[width=0.9\linewidth]{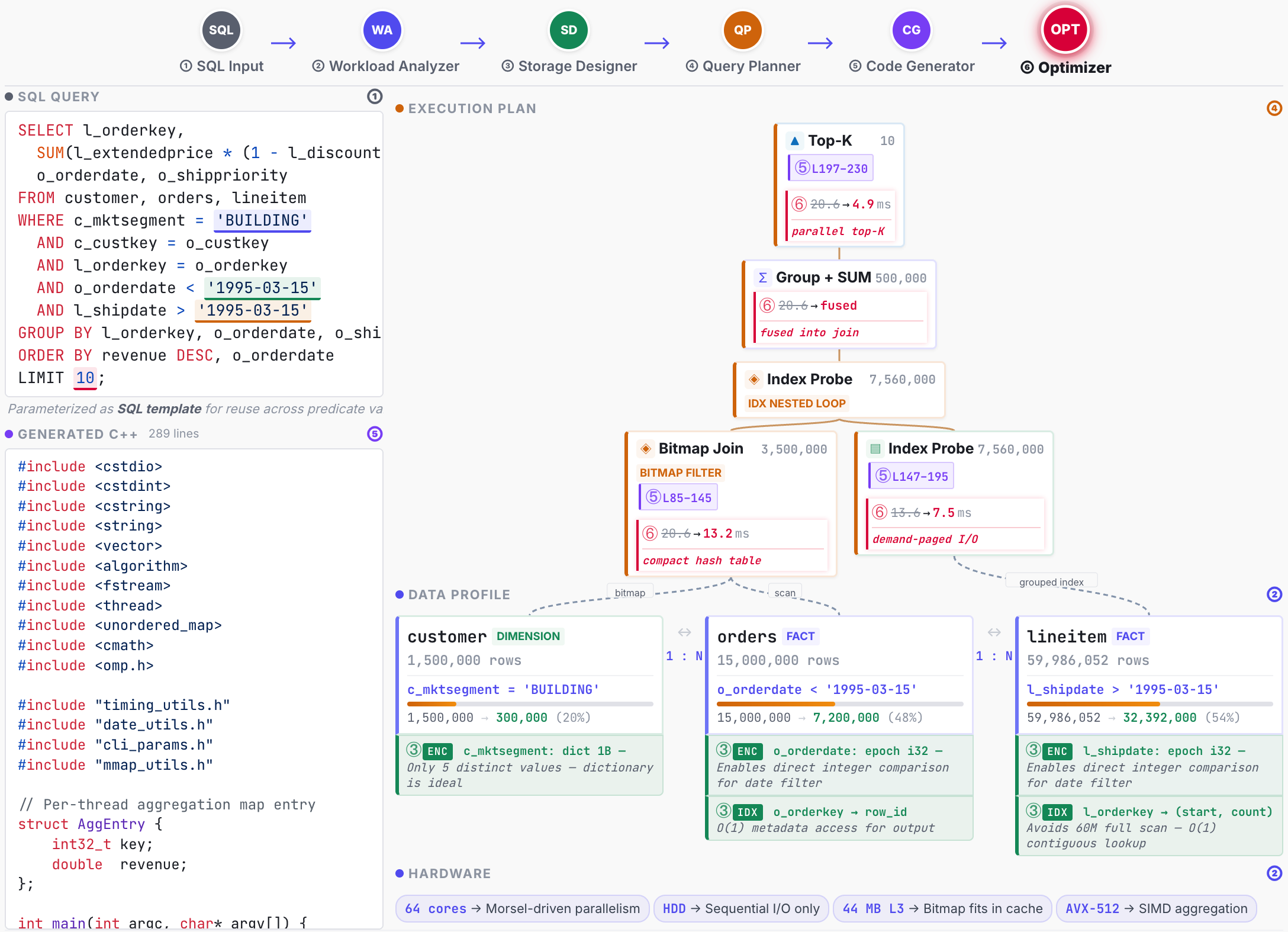}

    \caption{User Interface of GenDB. A live demo is available at {\color{blue} \url{https://solidlao.github.io/GenDB/demo/}}.}

    \label{fig:gendb_user_interface}
\end{figure*}

\noindent \textbf{Performance Comparison.} Figure~\ref{fig:perf_comparison} shows the results, and GenDB outperforms all baselines on both benchmarks. Note one reason is GenDB's optimization for offline code generation of repetitive, templated queries, whereas other systems do not exploit this offline advantage. On TPC-H, GenDB achieves a total execution time of {249 ms} across five representative queries~\cite{compile-vectorize}. This is {2.4$\times$} faster than DuckDB (596 ms), {2.7$\times$} faster than Umbra (660 ms), and {9.2$\times$} faster than ClickHouse. On SEC-EDGAR, GenDB achieves {403 ms}, which is {3.8$\times$} faster than DuckDB and {3.3$\times$} faster than Umbra. 
GenDB uses iterative optimization with early stopping criteria. On TPC-H, Q9 and Q18 reach competitive execution times of 105~ms and 45~ms respectively at iteration~0, requiring no further optimization.
Q3 starts at 593~ms and decreases to 27~ms by iteration~3, a {22$\times$} improvement. This gain comes from replacing a 1 GB max-orderkey-sized array with a compact 56 MB structure, switching to demand-paged I/O to eliminate unnecessary page faults, and parallelizing the dimension filter. 
Due to space limitations, full experimental results (e.g., all TPC-H queries) are available in our project repository at {\color{blue} \url{https://github.com/SolidLao/GenDB}}. We will prepare visualized results for all queries to facilitate discussion during the demo session.

\noindent \textbf{Cost Analysis.} We used Claude Sonnet 4.6 as the underlying LLM for GenDB. Synthesizing the engine cost \$14 and 91 minutes for TPC-H, and \$23 and 140 minutes for SEC-EDGAR. The higher cost and time for SEC-EDGAR are expected, since it is a new benchmark that LLMs must approach through trial and error. This also demonstrates that GenDB generalizes to benchmarks that LLMs are not familiar with, ultimately achieving good performance through iterative refinement. Moreover, this cost is acceptable for offline code generation, since the generated code template is executed frequently and repeatedly. For example, in 50\% of Amazon Redshift clusters, 80\% of queries exactly repeat previously observed queries, and long-running queries almost always recur across clusters~\cite{redset-vldb}. We plan to explore reducing generation cost by using smaller models for simpler steps and generating reusable operators.

\noindent \textbf{Correctness Analysis.} GenDB validates correctness by comparing execution output against ground-truth results from a traditional DBMS (e.g., DuckDB), and the generated code passes the fuzz testing, i.e., it returns consistent results across different SQL template predicates and testing databases. However, execution consistency does not guarantee true semantic equivalence: agreement across a set of test parameters and databases does not imply agreement across all possible databases. We plan to explore stronger correctness guarantees, such as formal verification, in the future.

\section{Demonstration Plan}

The demonstration setup includes a table with a laptop connected to a portable projector and a poster. The laptop is pre-configured to run GenDB to generate query processing code. While internet access to OpenAI GPT or Anthropic Claude is preferred, it is not required, as the demonstration includes cached results for convenience. The demonstration is structured to last five minutes, with additional details available upon request. Each visitor receives a brief introduction to explain the system and its web interface. The visitors can then proceed to use GenDB themselves.

Figure~\ref{fig:gendb_user_interface} shows the user interface, using TPC-H Q3 as an illustrative example. Note GenDB is not limited to pre-computed results. Users can write arbitrary SQL on any loaded benchmark and select different LLMs to generate optimized code in real time. Users can also upload their own datasets (via \texttt{CREATE TABLE} statements and CSV files) and queries; GenDB then generates storage layouts, execution plans, and optimized code tailored to the uploaded data. The top of the interface displays the multi-agent pipeline of GenDB as a clickable workflow bar; selecting any agent reveals its contribution to the integrated two-column visualization below. 
\circnum{1} The \textbf{SQL Input} step extracts parameterized SQL templates from queries and generates code at the template level. Queries that share the same template structure (e.g., different date ranges or filter predicates) reuse the same generated code.
\circnum{2} The \textbf{Workload Analyzer} profiles the target data and hardware to identify bottlenecks and optimization opportunities. Each table block on the canvas shows the cardinality, filter predicate, and selectivity bar---for example, the customer segment filter retains only 20\% of rows (1.5M~$\to$~300K). Hardware characteristics are mapped to execution strategies: 64~cores support morsel-driven parallelism, HDD favors sequential I/O, the 44~MB L3 cache can hold the 187~KB customer bitmap entirely, and AVX-512 is available for SIMD operations. Clicking this agent in the workflow bar shows further analysis, such as identifying the 60M-row \texttt{lineitem} table as the main bottleneck and noting that \texttt{c\_mktsegment} has only 5 distinct values, which is suitable for bitmap filtering.
\circnum{3} The \textbf{Storage Designer} transforms the raw text files into a binary columnar layout with memory-mapped I/O. Encoding and index badges appear on each table box with explanations: \texttt{c\_mktsegment} is dictionary-encoded into a single byte (5 distinct values), date columns use epoch-based \texttt{i32} for direct integer comparison, and a grouped index on \texttt{l\_orderkey} maps each order key to a contiguous \texttt{(start, count)} range, avoiding the 60M-row full scan with $O(1)$ lookup.
\circnum{4} The \textbf{Query Planner} generates the physical execution plan, shown as an operator tree in the right column with edges connecting operators to their source tables. For Q3, the plan first builds a bitmap from the 300K qualifying customers and applies it as a semi-join filter during the orders scan (3.5M surviving rows). It then probes the lineitem grouped index (7.5M rows), aggregates into 500K groups, and returns the top~10 by revenue.
\circnum{5} The \textbf{Code Generator} produces a C++ program (289 lines for Q3) that implements the plan. A syntax-highlighted code preview is shown in the left column below the SQL, and line-range badges on the operator nodes (e.g., L85--145 for the bitmap join) link each plan operator to its corresponding code section, allowing users to trace how logical operators map to the generated code.
\circnum{6} The \textbf{Optimizer} iteratively profiles the generated code and rewrites performance bottlenecks. Per-operator timing annotations on the plan tree show the effect of each optimization (e.g., the index probe improves from $13.6$ to $7.5$~ms through demand-paged I/O).
Clicking this agent shows the full optimization journey (omitted in the figure due to space constraints): for Q3, the initial code runs in 593~ms, with 91\% of the time spent zero-initializing a 1~GB array that stores only 3.5M entries. Over four iterations, the optimizer removes this overhead using compact data structures, switches to demand-paged \texttt{mmap}, and parallelizes the bitmap filter, reducing the runtime to 27~ms---a \textbf{22$\times$} improvement. Users can browse and compare the generated C++ code across iterations, with each version annotated by the bottleneck identified, the fix applied, and the resulting performance impact.




\section{Related Work}

GenDB relates to prior work that uses compilation techniques to generate customized code for efficient query processing~\cite{legobase-query-engine-scala-c, hique-holistic-code-generation-template}. HIQUE~\cite{hique-holistic-code-generation-template} generates customized code by replacing every operator in the execution plan with manually implemented C code templates. 
LegoBase~\cite{legobase-query-engine-scala-c} uses a high-level programming language, Scala, to implement query engines, and then employs a compiler to generate optimized C code for each query. These works are motivated by the observation that hand-written C/C++ code clearly outperforms even very fast vectorized systems~\cite{hyper-compilation-llvm}. 
Hiring human experts to write customized code for a large number of long-running queries is costly and does not scale. These methods were proposed as early attempts at automatic code generation. However, these methods provide limited customization because they rely on human-designed code templates. The generation process itself is fixed and unaware of data characteristics, workload patterns, or hardware properties.

GenDB also relates to work that uses LLMs for code generation~\cite{codexdb, trummer2025genesisdb}. CodexDB~\cite{codexdb} generates Python code for query processing by transforming each operator in an execution plan into a natural language description based on operator-specific text templates, and then invoking the GPT-3 Codex model to generate code. GenesisDB~\cite{trummer2025genesisdb} extends CodexDB by generating reusable implementations of general relational operators. These systems provide customization options, such as selecting the data processing library and enabling debugging features (e.g., printing intermediate results during execution). However, they do not target systematic performance optimization. Bespoke OLAP~\cite{wehrstein2026bespokeolapsynthesizingworkloadspecific} is concurrent work on query engine synthesis that focuses on exploiting workload characteristics, but unlike GenDB, it does not fully leverage hardware features (e.g., it supports only single-threaded code synthesis).



\section*{Acknowledgement}

This material is based upon work supported by the National Science Foundation under Award No.\ 2239326.

\bibliographystyle{ACM-Reference-Format}
\bibliography{sample}

\balance

\end{document}